# Reliable and redundant FPGA based read-out design in the ATLAS TileCal Demonstrator

Henrik Åkerstedt, Steffen Muschter, Gary Drake, *Member, IEEE*, Kelby Anderson, Christian Bohm, *Senior Member, IEEE*, Mark Oreglia and Fukun Tang
On behalf of the ATLAS TileCal collaboration

*Abstract–* The Tile Calorimeter at ATLAS [1] is a hadron calorimeter based on steel plates and scintillating tiles read out by PMTs. The current read-out system uses standard ADCs and custom ASICs to digitize and temporarily store the data on the detector. However, only a subset of the data is actually read out to the counting room.

The on-detector electronics will be replaced around 2023. To achieve the required reliability the upgraded system will be highly redundant. Here the ASICs will be replaced with Kintex-7 FPGAs from Xilinx. This, in addition to the use of multiple 10 Gbps optical read-out links, will allow a full read-out of all detector data.

Due to the higher radiation levels expected when the beam luminosity is increased, opportunities for repairs will be less frequent. The circuitry and firmware must therefore be designed for sufficiently high reliability using redundancy and radiation tolerant components.

Within a year, a hybrid demonstrator including the new read-out system will be installed in one slice of the ATLAS Tile Calorimeter. This will allow the proposed upgrade to be thoroughly evaluated well before the planned 2023 deployment in all slices, especially with regard to long term reliability.

Different firmware strategies alongside with their integration in the demonstrator are presented in the context of high reliability protection against hardware malfunction and radiation induced errors.

## I. Introduction

THE ATLAS Tile Calorimeter (TileCal) [2] is partitioned into 4 cylindrical sections, each composed of 64 wedge shaped modules. The electronics are located in extractable super-drawers at the base of the wedges. This places them at the outside of the calorimeter, shielded by the calorimeter iron tiles. The super-drawers contain either 45 or 32 PMTs depending on section, with each calorimeter cell read out by two PMTs. In total there are 9852 PMTs, reading out 4926 calorimeter cells. In the current system, the PMT signals are amplified and shaped by analog front-end boards, (3in1), with two different gains (with gain ratio 64). This makes it possible to cover a dynamic range of 16 bits using two 10 bit ADCs. The signals are digitized and stored in pipelined memories to await a corresponding level-1 trigger accept before being read out off the detector. The trigger decision is formed from analog signals from the front-end boards, summed to represent tower geometries and transferred to the Level-1 calorimeter trigger in the counting room outside the detector. For calibration purposes, the 3in1 front-end boards can also produce realistic charge injection pulses well as read out the integrated response from a circulating cesium source.

By 2023 the LHC accelerator will be upgraded [3] to an instantaneous luminosity up to 5 times the nominal value and an average luminosity of up to 10 times the nominal average. This implies more complex events overlaid with minimum-bias backgrounds, as well as elevated radiation exposure to the electronics. Significant technology advances since the initial TileCal electronics were built, now allow much higher readout bandwidths and radiation-tolerant designs using commercial components. In the upgraded Tile Calorimeter electronics planned for 2023, PMT data for all LHC bunch crossings will be read out to the off-detector electronics [4], allowing the Level-1 trigger selection to make use of cell-by-cell digital processing, in contrast to the current pre-summed analog towers. This will help reduce the effect of minimum bias pile-up on triggering, which is expected to be a considerable problem in high luminosity operations.

To improve reliability, the proposed electronics [5] will be made more modular with smaller independent modules and improved redundancy. In contrast to the current system where each module covers 45 PMTs, the new system will be segmented into modules covering 6 PMTs each, with each calorimeter cell read out by two independent modules. If one module fails the cell will still be read out, although with somewhat worse precision. If both modules fail, only 6 cells will be lost, corresponding to 0.06 % of the detector. Mechanically the super-drawers will be replaced by 4 mini-drawers, with each mini-drawer containing 12 PMTs read out by two modules. Their smaller size will improve access for service or replacement during maintenance periods. Repair and replacements of failed components parts may be performed by extracting the failing mini-drawer (for later repair off-detector) and directly inserting a well proven one, thus limiting radiation exposure to personnel. Two redundant 10 volt power bricks deliver the low voltage power to each module, each with the power capacity to drive both modules.

To gain experience with this design, a hybrid demonstrator project (Fig 1) is being assembled, with the goal of installing a prototype module in a Tile calorimeter slice in 2014 or alternatively during the next possible shutdown period. The demonstrator must be able to operate seamlessly with the present system so that it will not disrupt data taking, so analog tower sums must still be provided to the Level-1 Calorimeter trigger  This consideration has also driven the choice of front-end solution for the demonstrator.

The demonstrator on-detector electronics is housed in the mini-drawers, each containing 12 modified 3in1 front-end boards, a main board for digitization, a daughter board for communication and control, and a high voltage power supply. The three latter are designed as two independent identical parts, each servicing six PMTs. There is also an assembly of legacy analog summation boards to provide compatibility with the current Level-1 trigger.

Two modulator-based QSFP+ modules are used for communication to provide radiation tolerance [6] and redundancy, each conveying four 4.8 Gb/s down-links for clock synchronization and configuration, and four 10 Gbps up-links for data and monitoring. The logic is implemented in Kintex-7 FPGAs [7] with elaborate error mitigation techniques to be able to recover from radiation-induced errors.

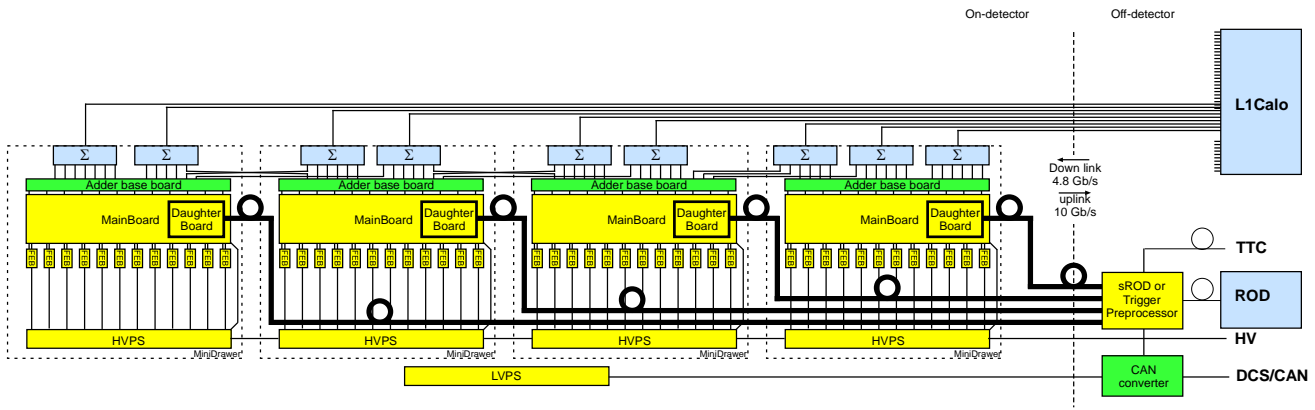

Fig 1. Overview of the demonstrator. The upgrade readout system (yellow) is augmented with legacy analog summing boards (blue) with adder base boards and a CAN converter (green) as interfaces compatible with the present trigger and DAQ

.

## II.  THE DAUGHTER BOARD

The Daughter Board [8] (Fig. 2), which is responsible for communication with the off-detector electronics, is being developed at Stockholm University. It is physically designed to reside on the main board designed at University of Chicago, which controls and digitizes data from the front-end 3in1 boards. To improve reliability, the system is designed to be inherently redundant, duplicating or triplicating all functions to minimize single point failure modes. The two halves of the board are designed symmetrically (Fig. 3), with each side serving 6 PMTs as mentioned above. There is a set of four 4.8 Gbps down-links, one of which goes to the CERN-developed radiation-hard protocol chip, the Giga Bit Transceiver (GBTx) [9]. The GBTx, in turn, connects to the JTAG chain to program the FPGAs. The other three down-links enter GTX transceivers in the FPGAs.  Connection with the main board is via a 400 pin high speed connector.

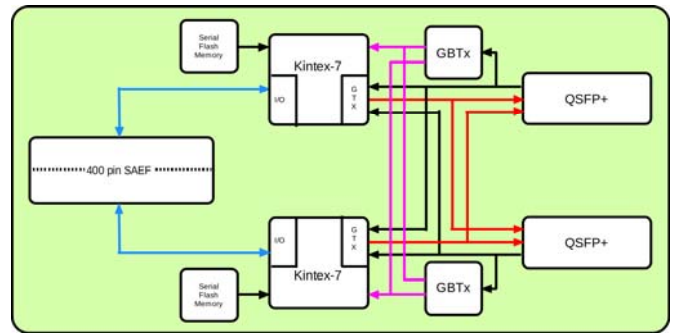

Fig. 3 Overview of the daughter board

Two identical Kintex-7 FPGAs are used for the logic functionality in the daughter board. A series of increasingly complex prototypes have been designed and a possible final version is currently being tested. The demonstrator must be thoroughly tested to verify functionality and radiation tolerance before being installed in order to ensure the data integrity. Radiation testing of separate components is currently underway.

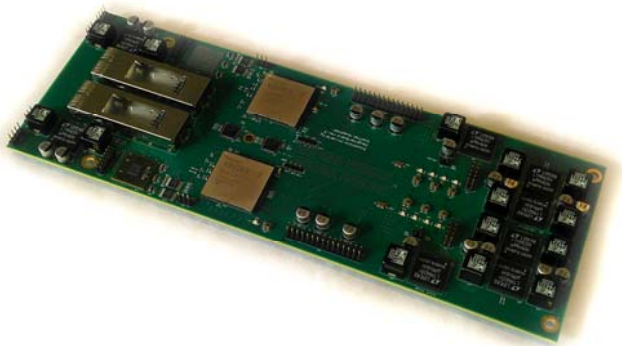

Fig. 2 The daughter board

## III. RADIATION INDUCED ERRORS IN FPGAS

To limit the effect of radiation induced damage as well as component failures, the overall firmware design must be carefully considered. The number of gates needed for a given design is much higher in FPGAs than in an ASIC. An FPGA is therefore more susceptible to radiation damage. Radiation affects electronics cumulatively as well as on a single event basis. With the small feature sizes of current IC technologies, the risk of permanent faults is reduced, but the smaller feature sizes also means smaller charges are needed to upset logic states of flip/flops. Single event upsets (SEUs) are thus more frequent. However, SEUs can be detected and corrected using firmware activated mechanisms that have been incorporated in the FPGA design.

## IV. MITIGATION OF RADIATION INDUCED ERRORS

To a large extent, SEUs can be mitigated in the FPGA firmware, but at the expense of increasing the gate count. SEUs are treated differently depending on if the errors appear in the configuration memory or in the logic fabric. Errors in the configuration memory can be dealt with using scrubbing or partial reconfiguration. Scrubbing repeatedly reads all configuration words, corrects single bit errors and adjacent double bit errors. Other multi-bit configuration errors are reported when detected by the scrubber. These errors can be mitigated by full or partial reconfiguration. Single and multiple upsets in the logic fabric as well as long term damage can be mitigated by redundant data paths and voting (Triple Module Redundancy, TMR). If three such modules are well separated geographically the likelihood that a MEU will affect more than one module is small, which means that the error will not cause any data failure. During the duration of a failure (until overwritten by new data or fixed) the TMR function is inactive. A new failure in the remaining two modules will not be corrected. This interval should be minimized to minimize the probability for uncorrected errors.

All important functions of the design will be protected using TMR. Data integrity is ensured with check sums in the transmission links. Local memories are avoided when possible, but will have to be protected by error correction mechanisms. Fast detection, good coverage and effective repairs are crucial to reach the reliability level that is necessary for this project.

## V. RADIATION TEST RESULTS

Radiation testing has so far focused on SEUs. Accelerated life testing of neutron damage is scheduled for summer 2014, along with other components of the demonstrator.

The SEU tests of the daughter board have been performed at Massachusetts General Hospital (MGH) Francis H. Burr Proton Therapy Center during this year. Their cyclotron delivers a very versatile proton beam with a wide range in energy and flux.

Running at a flux of $8 \cdot 10^4$ protons/cm$^2$/s (216 MeV) for approximately one hour resulted in an equivalent 100 day run at a nominal luminosity of $10^{34}$cm$^{-2}$s$^{-1}$. The data gathered from the device under test was used to estimate the expected rate of SEUs, the scrubber success rate, and to give a lower limit to the number of transmission errors in the FPGAs transceivers. The results achieved are compatible with results from test [10] performed by the ATLAS Liquid Argon collaboration.

For a given Kintex FPGA in a daughterboard under post upgrade luminosity, we expect 30 one-bit errors per week and 1 repairable two-bit errors per week. Unrecoverable errors will be less frequent, but will nonetheless happen 1-2 times every month. This will in most cases require reconfiguration, amounting to a maximum of one minute of downtime and partial loss of statistics in one mini-drawer. However, with partial reconfiguration TMR should significantly reduce the consequences of the unrecoverable errors.

Transmission errors, caused by upset flip-flops in the design will happen 5 times per month. However with redundant links transmitting CRC encoded data, this will most likely not cause any loss in data.

## VI. SUMMARY

The use of FPGAs in the Tile calorimeter gives us an unprecedented level of flexibility, but it also means that the design is more susceptible to radiation induced SEUs.

With the relatively low local flux after the luminosity increase and with the right mitigation strategies, we expect the reliability to reach acceptable errors even considering the fact that we plan to use ~2000 FPGAs. However, more investigations are needed to reach this point.

## VII. ACKNOWLEDGEMENT

## REFERENCES

[1] ATLAS collaboration, "The ATLAS Experiment at the CERN Large Hadron Collider", 2008 JINST 3 S08003.

[2] ATLAS collaboration, "ATLAS Tile Calorimeter: Technical Design Report", CERN-LHCC-96-042 (1996).

[3] ATLAS collaboration, "Letter of Intent for the Phase-II Upgrade of the ATLAS Experiment", CERN-2012-022, 2012

[4] sROD

[5] F. Tang et al., "Design of the front-end readout electronics for the ATLAS tile calorimeter at the sLHC," IEE Transactions on Nuclear Science, Vol. 60, NO. 2, 1255-1259, April. 2013.

[6] G Drake et al., "A new high-speed optical transceiver for data transmission at the LHC experiments", 2014 JINST 9 C01059

[7] XILINX, 7 Series FPGAs Overview, DS 180, November 2012.

[8] S. Muschter et al. "Development of a readout link board for the demonstrator of the ATLAS tile calorimeter upgrade".

Topical Workshop on Electronics for Particle Physics 2012, JINST 8 C03025.

[9] P. Moreira et al., "The GBT project", in the proceedings of the Topical Workshop on Electronics for Particle Physics (TWEPP 2009), September 21–25, Paris, France (2009), CERN-2009-006.

[10] M J Wirthlin., "Soft error rate estimations of the Kintex-7 FPGA within the ATLAS Liquid Argon (LAr) Calorimeter", 2014 JINST 9 C01025